# A Survey Instrument to Assess Students' AI and Generative AI Knowledge

Aditya Johri
*Information Sciences & Technology*
*George Mason University*
Fairfax, USA
johri@gmu.edu

Cory Brozina
*Ryan School of Engineering, College of STEM*
*Youngstown State University*
Youngstown, USA
scbrozina@ysu.edu

Akriti Bagale
*Information Sciences & Techology*
*George Mason University*
Fairfax, USA
abagale@gmu.edu

***Abstract*— In this research-to-practice paper we present a survey that can be used to assess students' AI knowledge. As the use of artificial intelligence (AI), including generative artificial intelligence (GenAI), has proliferated, so has the need to educate students about the topic. A range of AI literacy frameworks have been proposed, outlining the essential knowledge that students should have. Alongside, different ways of assessing AI knowledge have been developed. As yet, there is a lack of assessment instruments capable of evaluating multiple forms of student knowledge, including technical concepts, practical applications, and ethical concerns about AI use. In this article, we present a study implementing a comprehensive instrument to assess AI knowledge. The instrument combines measures from multiple scales to capture a range of literacy features and actual knowledge. We implemented the instrument in a higher education setting to assess its viability and usefulness and found that the instrument exhibited useful diagnostic capabilities and was able to identify common misconceptions among students. Although students performed well overall, there was a significant misunderstanding of how AI, especially GenAI systems, work. It also identified a lack of higher-level knowledge. The instrument is publicly available for use by others. We foresee its usefulness as a diagnostic that goes beyond understanding students' attitudes and perceptions of AI and GenAI use and tests multiple aspects of students' knowledge and conceptual understanding. This can enable the development of targeted instruction.**

## I. Introduction

As the use of AI, including GenAI, has proliferated, so has the need to educate both users and future developers on the technology. Within education, the use of AI is experiencing exponential growth [1], and across all facets of education, from personalized learning to assessment, AI applications are being developed and deployed at scale. Furthermore, the introduction of conversational large language models (LLMs) based applications such as ChatGPT have completely upended many aspects of teaching and learning, especially in higher education institutions (HEIs). HEIs are releasing a range of guidelines for how GenAI should be used but at the same time it has also become important for them to increase the opportunity students have to learn about AI. Overall, across HEIs, there is an urgent need to both provide AI literacy and assess students' AI knowledge to better understand what they know and which interventions and training are required to improve their AI expertise.

In this paper, we present a study that evaluates a survey to assess students' AI literacy. Although many ways of assessing AI literacy have been proposed in the literature, as we review below, most are exploratory [2], and, with a few exceptions, lack items that provide a comprehensive evaluation [3]. While many forms of knowledge assessment are available, and many ways of testing knowledge, such as performance on hands-on activities, are more situationally appropriate, a short survey remains one of the most useful tools we have for this purpose. The survey we have curated using multiple sources examines students on a range of items that cover all elements of AI literacy, including technical knowledge, understanding of AI use, and understanding of ethical and responsible aspects of AI. In the remainder of the paper, we briefly discuss prior work, the curation of the instrument, and then present the results of the diagnostic study. We conclude with a discussion of our findings.

## II. Relevant Prior Work

### *A. AI Literacy Constructs*

The first step to be able to ascertain AI literacy among students is to establish a criteria for what is AI literacy. Like other forms of literacy that were studied previously, such as media literacy, digital literacy, and even data literacy, there is a need for consensus around constructs contained within the larger concept. As research on AI literacy has developed, certain elements of the concept have started to appear consistently within the literature. Currently, the following broad constructs have found significant acceptance as being core to AI literacy in terms of students' knowledge development [4, 5]:

1) *Know and Understand*: Knowledge and understanding of AI fundamental concepts and techniques is a common construct across studies and refers to knowing the basics of AI and under-standing how AI applications work from input data to machine learning techniques. It also includes understanding how data is created including through sensors, and to know that humans play a significant role in the development of AI.

2) *Use and Apply*: In relation to AI, this construct refers to the operational or functional aspects of AI use. In particular, the ability to use AI tools to accomplish tasks. There is a wide range of uses and applications related to AI that can be put into practice depending on the needs of the user. These can be for work or school, everyday activities such as listening to music, watching movies, or finding using a map-based application.

3) *Evaluate and Create*: Evaluation involves the ability to analyze and interpret the outcomes of AI applications critically. Having a comprehensive understanding of the technical aspects of AI enables individuals to examine and form informed opinions about their interactions with AI technologies. Create is a less common construct within the literature, but it emphasizes the ability to design and code AI applications. Ng et al. [2] combined “evaluate and create” into a single construct representing higher-order thinking skills, and we use that variation, especially since the student population we surveyed was unlikely to create high-end AI applications.

4) *Navigate Ethically*: Understanding the ethical and societal implications of AI is crucial to educating citizens to become socially responsible and ethical users of AI. Human-centered considerations such as fairness, accountability, and transparency must be prioritized in addition to AI aspects such as privacy, misinformation, and bias. An AI-literate person must be able to understand and judge ethical issues to ensure that the use and development of future AI technology are grounded in principles such as inclusivity, equitable access, and minimizing potential bias.

### B. *AI Literacy Assessments*

The constructs discussed above have guided the design and implementation of a range of pedagogical interventions across K12 [6] and higher education [7]. These include both formal and out-of-school implementations of courses, curricula, and hands-on activities. These interventions have been assessed using different assignments and instruments. One of the first assessment instruments was developed by Wang and colleagues using a scientifically rigorous psychometric scale to measure AI literacy [8,9]. The scale consists of 12 items that cover four core constructs: “Recognize,” “Use and Apply,” “Evaluate,” and “Ethics.” The overall scale has satisfactory convergence validity, but using distinct constructs alone has yet to yield reliable results. [10] presented a comprehensive “Meta AI Literacy Scale-MAILS,” a subjective assessment scale with 34 items. Another general instrument intended for non-experts across disciplines is the assessment tool developed by [11]. It consists of 38 items, was drawn from prior work, and was validated using the Delphi method. The researchers found that affective-related items, such as "attitudes towards AI," were not included in the final set of items, indicating that they are not essential to AI literacy. [12] developed another scale to measure employees’ AI literacy. Their instrument consisted of 13 items and was based on a human-AI interaction theory.

Although there are a range of instruments, they mostly study attitudes or experiences related to AI, or the efficacy of the intervention itself, but not factual knowledge. For this study, we were interested in factual knowledge that is aligned with the constructs. After further research into secondary sources, we identified three instruments, discussed below. We curated items from these and modified them slightly to suit our purpose. Our contribution here is not unique in terms of development of items from scratch but in terms of curating them from different instruments in order to develop a survey that was efficient in terms of time required for responding; captured all elements for AI literacy; and measured actual knowledge of students. Furthermore, the curated survey is broader in scope, covers both basic and advanced knowledge, and covers both AI and GenAI knowledge while the other surveys focus on one or the other.

## III. Curating the AI Literacy Assessment Instrument

### A. *GLAT Instrument*

The GenAI Literacy Assessment Test (GLAT) is a 20-item multiple-choice instrument that tests both AI and GenAI knowledge [13]. This instrument was developed precisely for the drawback identified above with existing instruments in that they largely rely on self-reports of attitudes and experiences, which may be biased, rather than objective tests of knowledge. GLAT was developed using established procedures in psychological and educational measurement, and its structural validity and reliability were assessed and confirmed using data from 355 higher education students. The testing showed that GLAT scores significantly predicted learners’ performance in GenAI-supported tasks and that it outperformed self-reported measures of perceived proficiency.

We used most of the items in the final GLAT instrument with a few modifications. Many items in the GLAT instruments, even though the title of the test is GenAI Literacy Assessment, are broader in scope and about AI literacy more generally. We used these items but moved them to the part of the survey focused on AI and not GenAI, which was a subset of our larger survey. We also created items similar to GLAT to address AI more broadly (e.g., “Which of the following best describes AI”, Item 1).

### B. *Pew Survey*

The Pew AI literacy questionnaire was developed by the Pew Research Center in consultation with Ipsos [14]. The web program was rigorously tested on both PC and mobile devices by the Ipsos project management team and Pew Research Center researchers. The Ipsos project management team also populated test data, which was analyzed in SPSS, to ensure the logic and randomizations were functioning as intended before launching the survey. The quiz consists of six items, all of which were used in our survey. Taken together, 30% of Americans answered all six questions correctly about AI awareness in everyday life (high awareness), 38% answered three to five questions correctly (medium awareness), and 31% answered two or fewer questions correctly (low awareness). The mean number of correct answers was 3.7 out of 6.

We used this survey instrument, given its validated testing, development, and implementation, as a baseline that students would at least know or should know what the general U.S. population knows about AI. When using items from the Pew survey, we modified the following two items. 1. When playing music, which of the following uses artificial intelligence (AI)? Modified item: You are a music enthusiast and are evaluating the use of AI to improve your listening experience. For which of these activities will you use artificial intelligence (AI)? 2. Thinking about devices in the home, which of the following uses artificial intelligence (AI)? Modified item: As a technology enthusiast, you are always looking to make your home “smarter” by using different devices and applications. For which of these activities, will you use your knowledge of artificial intelligence (AI)?

**TABLE I: ITEMS INCLUDED IN THE SURVEY**

| AI QUESTIONS | | |
|---|---|---|
| *Know and Understand* | | |
| 1 | Which of the following best describes Artificial Intelligence (AI). | a. Artificial intelligence (AI) refers to a set of computer science activities dedicated to the process of creating, designing, deploying and supporting software.<br>b. Artificial intelligence (AI) is technology development methodology that follows an iterative and incremental approach to software development to deliver a working product quickly and frequently.<br>c. Artificial intelligence (AI) is technology that enables computers and machines to simulate human learning, comprehension, problem solving, decision making, creativity and autonomy.<br>d. Artificial intelligence (AI) is the process of creating a plan for solving a problem by defining a computational process that meets system specifications. |
| 2* | Thinking about customer service, which of the following uses artificial intelligence (AI)? | a. A detailed Frequently Asked Questions webpage<br>b. An online survey sent to customers that allows them to provide feedback<br>c. A contact page with a form available to customers to provide feedback<br>d. A chatbot that immediately answers customer questions |
| 3* | When using email, which of the following uses artificial intelligence (AI)? | a. The email service marking an email as read after the user opens it<br>b. The email service allowing the user to schedule an email to send at a specific time in the future<br>c. The email service categorizing an email as spam<br>d. The email service sorting emails by time and date |
| 4* | Thinking about health products, which of the following uses artificial intelligence (AI)? | a. Wearable fitness trackers that analyze exercise and sleeping patterns<br>b. Thermometers that are placed under someone's tongue to detect a fever<br>c. At-home COVID-19 tests<br>d. Pulse oximeters that measure a person's oxygen level of the blood |
| 5 | In the context of AI, what is "zero-shot learning"? | a. Training a model without any data.<br>b. The ability of a model to perform a task without any task-specific training.<br>c. A method of reducing the model's training time to zero.<br>d. A technique for generating synthetic training data. |
| 6* | Thinking about online shopping, which of the following uses artificial intelligence (AI)? | a. Storage of account information, such as shipping addresses<br>b. Records of previous purchases<br>c. Product recommendations based on previous purchases<br>d. Product reviews from other customers |
| *Use and Apply* | | |
| 7* | You are a music enthusiast and are evaluating the use of AI to improve your listening experience. For which of these activities will you use artificial intelligence (AI)? | a. Using Bluetooth to connect to wireless speakers<br>b. Creating a playlist recommendation<br>c. Building a wireless internet connection to stream the music<br>d. Using shuffle mode for a chosen playlist |
| 8* | As a technology enthusiast, you are always looking to make your home "smarter" by using different devices and applications. For which of these activities, will you use your knowledge of artificial intelligence (AI)? | a. Programming a home thermostat to change temperatures at certain times<br>b. Programming a security camera that sends an alert when there is an unrecognized person at the door<br>c. Programming a timer to control when lights in a home turn on and off<br>d. Programming an indicator light that turns red when a water filter needs to be replaced |
| 9 | After deploying a customer service chatbot, you notice that it frequently provides outdated information about company policies. What is the best course of action to address this issue? | a. Implement a feedback loop where users can flag outdated information for review.<br>b. Schedule regular updates to the chatbot's training data to include the latest company policies.<br>c. Set up a system where complex or policy-related queries are escalated to human agents for accurate responses.<br>d. Conduct a comprehensive audit of the chatbot's performance metrics to identify areas for improvement. |
| *Evaluate and Create* | | |
| 10 | While reviewing a video of a well-known public figure making controversial statements, which characteristic confirms the video was NOT generated by AI? | a. The public figure's voice sounds like themselves.<br>b. The video has a professional and polished appearance.<br>c. The video is high-quality with smooth transitions.<br>d. None of the above. |
| 11** | You are building a computer vision application and evaluating different sources of data for your product. Which of the following | a. Infrared sensor<br>b. Speaker<br>c. Mobile phone camera<br>d. Medical Compound Tomography (CT) scanner |
| | devices cannot provide you with relevant data? | |
| 12** | You are building an AI system but its performance is under par. You are looking at ways to improve the output quality. Which of the following will most likely enhance an AI system's performance? | a. Increasing the learning rate<br>b. More noise in input training data<br>c. Using fictitious data for training<br>d. Larger amount of input training data |
| 13** | As part of a course on AI, you have learned about Natural Language Processing (NLP) techniques you have tasked with creating a tool that uses NLP. Which one of these is NOT a tool that you will create? | a. Clinical text record analysis<br>b. Gesture-controlled robot<br>c. Story generator<br>d. Question and answering tool |
| *Ethics* | | |
| 14** | Which of the following actions of an AI developer is NOT ethical? | a. Avoid providing details of the AI application to cover up the limitations of their products and services<br>b. Cross-checking the training dataset to ensure it is balances<br>c. Data collected is only use of training and testing AI applications<br>c. Ensure their products and services will not cause any foreseeable or unintentional harm |
| 15 | In a healthcare startup, an accurate AI model recommends treatments, but doctors don't trust it because they can't understand how the model arrived at its conclusions. What core issue does this scenario illustrate? | a. The AI model uses obsolete training data.<br>b. The training dataset lacks sufficient diversity.<br>c. The treatment guidelines input are incorrect.<br>d. The AI model behaves as a black box. |
| | **GENAI QUESTIONS** | |
| *Know & Understand* | | |
| 1 | Which of the following best describes "Generative AI"? | a. AI that creates new content like text, images, or music by learning from existing data.<br>b. An AI system designed to enhance the speed and accuracy of data retrieval in search engines.<br>c. A form of artificial intelligence that focuses on translating languages in real-time.<br>d. AI technology used primarily for managing and organizing large databases. |
| 2 | Which of the following statements best describes an LLM (Large Language Model)? | a. It generates text by analyzing and summarizing large volumes of web content.<br>b. It generates text by predicting the next word based on the context of previous words.<br>c. It generates text by translating input text into multiple languages simultaneously.<br>d. It generates text by using pre-defined templates and filling in the blanks. |
| 3 | Which of the following tasks can Generative AI perform with a high degree of accuracy? | a. Predicting stock market trends<br>b. Making ethical decisions in complex scenarios<br>c. Diagnosing rare diseases<br>d. Generating human-like text based on prompts |
| 4 | Which of the following is a potential challenge when using prompt-based development for text generation? | a. The language model can only generate binary outputs.<br>b. The need for extensive labeled data to train the model.<br>c. Crafting a prompt that accurately captures the desired context and nuances.<br>d. The requirement for complex feature engineering. |
| 5 | What does the term "token" refer to in the context of a large language model (LLM)? | a. A token is a unit of text, such as a word or a subword, that the model processes individually.<br>b. A token is a unique identifier assigned to each user interacting with the language model.<br>c. A token is a security measure used to authenticate API requests to the language model.<br>d. A token is a reward given to users for contributing valuable data to train the language model. |
| 6 | How does RAG (Retrieval-Augmented Generation) enhance the capabilities of an LLM? | a. By improving its grammar and syntax.<br>b. By providing it with real-time and relevant data.<br>c. By increasing its computational speed.<br>d. By enabling it to understand multiple languages. |
| *Use & Apply* | | |
| 7 | When using generative AI to create a marketing pitch, which of the following strategies is least likely to be effective? | a. Supplying the AI with information about the target audience<br>b. Asking the AI to include unique selling points and benefits<br>c. Requesting the AI to use persuasive language techniques<br>d. Providing the AI with a list of competitors' products |

| 8 | Suppose you have a large dataset of emails and you want to build an application to answer questions based on this dataset. Which of the following scenarios best illustrates the advantage of using RAG over prompting (i.e., without RAG)? | a. You need to generate creative writing pieces based on the email content.<br>b. You want to ensure the model can answer questions even if it has never seen similar questions before.<br>c. You need to answer questions that require specific information from different parts of the email dataset.<br>d. You want to reduce the size of the language model to save computational resources. |
|---|---|---|
| *Evaluate & Create* | | |
| 9 | As a student using a Large Language Model (LLM) to gather information for an assignment, how should you approach the information it provides? | a. The LLM's answers are always more trustworthy than any information you will find on the internet, so you can use them without further verification.<br>b. The LLM's answers are generally more trustworthy than internet sources, but you should still verify the information with other reliable sources.<br>c. The LLM's answers are not necessarily more trustworthy than internet sources, and you should cross-check the information with other credible references.<br>d. The LLM's answers are less trustworthy than internet sources because it relies on outdated information. |
| 10 | It is unlikely for an LLM to provide an accurate summary of the latest financial market trends in real-time. Is this statement true or false? | a. True, because the LLM's data may be outdated due to its knowledge cutoff.<br>b. True, because the LLM is not good at handling numbers and structured data.<br>c. False, because the LLM frequently updates its knowledge base.<br>d. False, because the LLM is capable of synthesizing the latest market data automatically. |
| 11 | A generative AI tool has provided a summary of a research paper. The summary states, "The study found that increased screen time is directly correlated with decreased attention spans in children aged 8-12." What is your next step? | a. Accept the summary as accurate because AI tools are generally reliable.<br>b. Ask the AI to provide more details about the study's methodology and results.<br>c. Cross-check the summary with the original research paper.<br>d. Use another AI tool to generate a summary for comparison and evaluate the consistency between both summaries |
| *Ethics* | | |
| 12 | What are the potential copyright implications for a journalist using an AI-generated image in a commercial article? | a. The AI-generated image is automatically free to use without any restrictions.<br>b. The journalist must pay a standard licensing fee to use the AI-generated image.<br>c. The image cannot be used in any commercial context because it is AI generated.<br>d. The journalist needs to check the licensing policy of the AI tool they used. |
| 13 | Should we impose restrictions on the outputs of generative AI technologies? | a. Yes, to reduce the computational resources required for operating these technologies.<br>b. Yes, to prevent the dissemination of harmful or misleading content.<br>c. No, as it would hinder technological innovation and creativity.<br>d. No, because users should have the freedom to access all generated content. |
| 14 | Sending personal information to cloud-based generative AI tools has little privacy concerns. | a. True, as this information is encrypted using sophisticated algorithms during the transmission process.<br>b. True, as generative AI tools are black-box systems and cannot output personal information even if it is used for model training.<br>c. False, as generative AI tools train on unencrypted data and can output private information based on their probabilistic nature.<br>d. False, as advancements in quantum computing can easily decipher the encrypted data. |
| 15 | When a generative AI system is used for screening job applications, what issue might arise concerning the quality and fairness of hiring decisions? | a. The AI system might overlook applicants' unique achievements and extracurricular activities.<br>b. The AI system could misinterpret minor formatting differences in resumes.<br>c. The AI system might not effectively handle applications submitted in various languages.<br>d. The AI system could reinforce existing biases found in historical hiring data. |
| **Item (Source: Jin et al., (2025), *except* 6 items marked* based on Pew (2023), and 4 items marked** based on Chiu et al (2024)** | | |

### C. *Chiu et al. (2024)*

[15] was one of the first studies to design a questionnaire to accurately assess AI literacy. They recognized that principally AI literacy studies were using self-reported questionnaires that assessed students' perceived AI capability rather than actual AI knowledge, as they asked learners about their self-perceptions. They wanted to develop something akin to other assessment instruments that targeted objective measures, such as those used in science, mathematics, and, more recently, for digital and computational literacy. They developed and validated an AI literacy test for school students comprising 25 multiple-choice questions. This instrument was validated as part of an AI curriculum for middle schools. 2390 students in grades 7 to 9 took the test, and it was validated using a Rasch model. The validation results showed that the model, and hence the instrument, met the dimensionality criteria, and therefore the items were both reliable and valid.

Table 1 above lists all the survey items and identifies the parent instrument for each.

## IV. Implementation of the Assessment Instrument

### A. Course and Student Population

The AI literacy survey was administered in an undergraduate Information Technology course taken by students who are juniors and seniors, i.e., 3rd and 4th year of their programs. A majority of students were minoring in cybersecurity, while the rest were pursuing concentrations in networking, web development, or cloud computing. We wanted to have some homogeneity among the student population in terms of their expertise and knowledge since this was the first implementation of this survey. We focus on students in information technology as the knowledge of AI is essential for them in terms of career preparedness and some level of proficiency with AI could be assumed for the population. Our thinking behind this student population as the first target for this survey was that if this format of the survey – which is more generic in nature – works for this student population, it can be further modified with additional domain or discipline related items for other students.

The course where the survey was administered covered sociotechnical and ethical aspects of technology. Topics included privacy and surveillance, sustainability, and economic implications of technology. Within the course, there was no specific lecture on AI, but students were exposed to the topic through different content in the course, including readings about AI and case studies on the use of AI in agriculture and education. At the time the students took this course, they had also not taken any other AI-related courses. This was by design; the survey was created to better understand their baseline knowledge so that future AI courses could teach them relevant concepts. In other words, the survey was designed as a diagnostic or formative assessment tool rather than a summative assessment.

### B. Survey Administration

The survey was administered online as an extra-credit assignment and implemented using the institution's standard examination system. This online exam system consisted of a lockdown browser functionality – called Respondus – and video monitoring capability to restrict cheating or plagiarism. We used this functionality to ensure that students completing the survey did not receive outside help and that their responses reflected their true knowledge. The average time needed to complete the survey was 13 minutes. Incomplete responses were removed from the sample, and the final sample size for analysis was 57 complete responses out of a total student population of 80. The study was approved by the institutional research board (IRB).

## V. Findings

In this section, we present the aggregate findings of the student survey, followed by a detailed breakdown and analysis for each assessment category.

### A. Aggregate Findings

The average score on the 30 survey items was 78.83% with a Cronbach's alpha of 0.66 and a 95% confidence interval of [.516,.775]. Item-level performance ranged from 100% on item Q6 (highest) to 33.33% on item Q17 (lowest).

By AI literacy category (Fig. 1), the average scores were: Know & Understand: 81.73% (SD=28.73), Ethics: 80.99% (SD=32.52), Evaluate & Create: 74.94% (SD=41.70), and Use & Apply: 74.74% (SD=38.05). The scores on individual questions within each category are shown in Fig. 2.

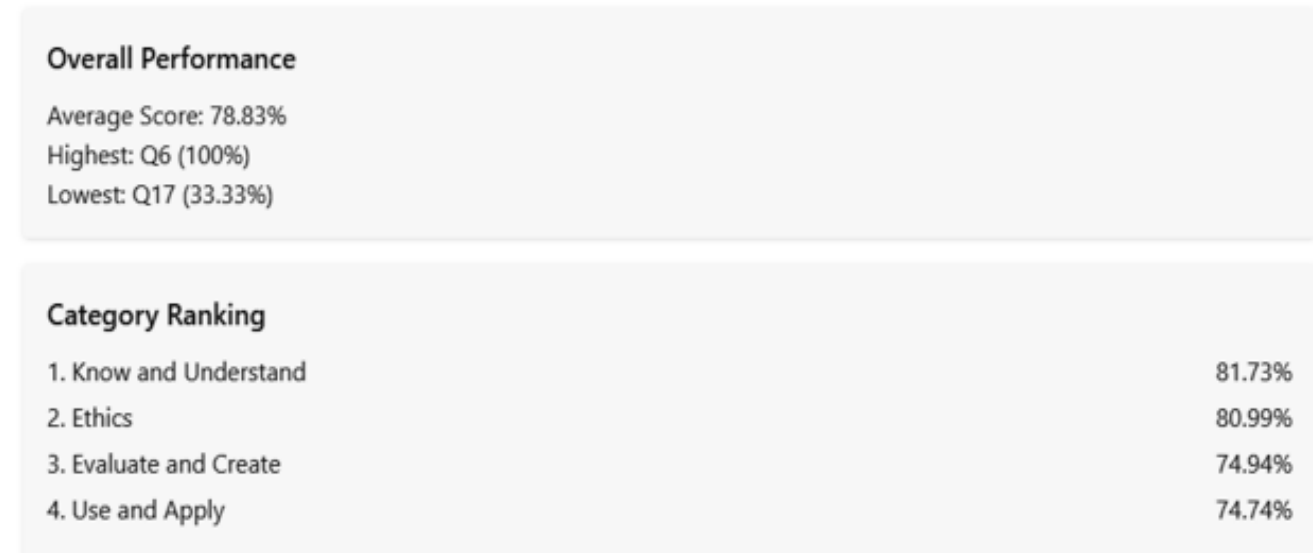

Fig 1 : Average score by AI literacy category

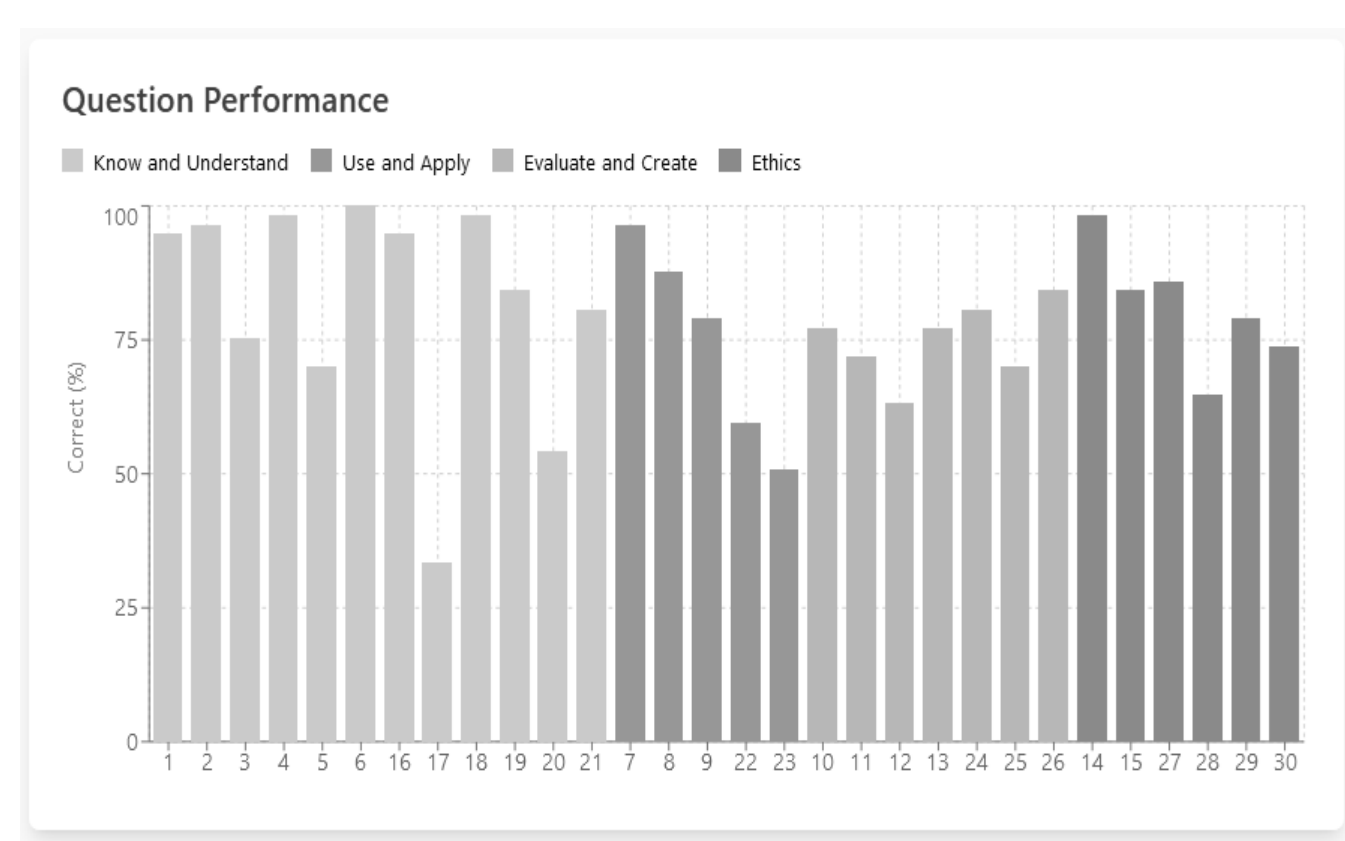

Fig 2: Average score by individual questions

Item difficulty was skewed toward higher performance (Fig. 3), but there was a reasonable spread: 8/30 items >90%, 7/30 between 80–90%, 9/30 between 70–80%, 2/30 between 60–70%, and 4/30 <60%. These distributions, together with the category average, indicate strong foundational knowledge and ethical reasoning, contrasting with comparatively lower performance on application (Use & Apply) and higher-order tasks (Evaluate & Create).

The most informative questions missed are conceptual mechanisms and tool-task alignment. On Q17 (LLM mechanism), only 33.33% identified next-token prediction, with a majority choosing the "LLMs summarize the web" misconception. Similarly, on Q23 (RAG vs. Prompting), just 50.88% recognized RAG's advantage as retrieving specific evidence from a corpus, with 36.84% overgeneralizing RAG as improving generalization to unseen questions.

Distribution Details

| Score Range | Questions | Percentage |
|---|---|---|
| >90% | 8 | 26.7% |
| 80-90% | 7 | 23.3% |
| 70-80% | 9 | 30.0% |
| 60-70% | 2 | 6.7% |
| <60% | 4 | 13.3% |

Fig 3: Item difficulty spread across all questions

For applied prompting (Q22), 59.65% correctly identified that supplying a competitor list is the least effective strategy, though 22.81% undervalued structured persuasive prompting.

Overall, more than 80% of the survey respondents scored above 70%, with 19.30% scoring 90% or higher (Table 2). Twenty students (35.09%) scored in the 80% range, while 29.82% scored in the 70% range. The vast majority of respondents scoring well on the survey lends support for using it for the associated course.

TABLE 2

| Score Range | Count | % |
|---|---|---|
| 0.90-1.00 | 11 | 19.30% |
| 0.8-0.899 | 20 | 35.09% |
| 0.7-0.799 | 17 | 29.82% |
| 0.6-0.699 | 5 | 8.77% |
| 0.5-0.599 | 3 | 5.26% |
| <0.5 | 1 | 1.75% |

Since students overall did reasonably well on the survey, we dove deeper to identify common misconceptions. We labeled questions that had a correct response rate of 75% or lower as misconceptions. Of the 30 questions, we categorized 11 questions as misconceptions. We will focus the rest of our analysis on these questions, which range in scale from 33.33% correct (question 17) to 75.44% correct (question 43).

## B. *Know and Understand*

Performance on the Know & Understand domain was generally high (M = 81.73%; 12 items), indicating a solid grasp of foundational concepts. The main exception was Q17, where only 33.33% answered correctly, leading to a significant dip in the distribution and revealing a specific misconception about how LLMs work (confusing next-token prediction with “summarizing the web”). In contrast, Q6 was answered correctly 100% of the time, indicating widespread mastery of that concept. Taking both into account, it shows that there is an overall strong knowledge with a specific gap. Participants reliably recognize core facts but have misconceptions in the underlying mechanism, which disproportionately depresses the domain’s meaning. This aligns with the study’s broader profile of stronger performance on foundational knowledge and ethics, with weaker spots in mechanism-level understanding and tool–task alignment, highlighting the need for brief, focused instruction to clarify the LLM's prediction process.

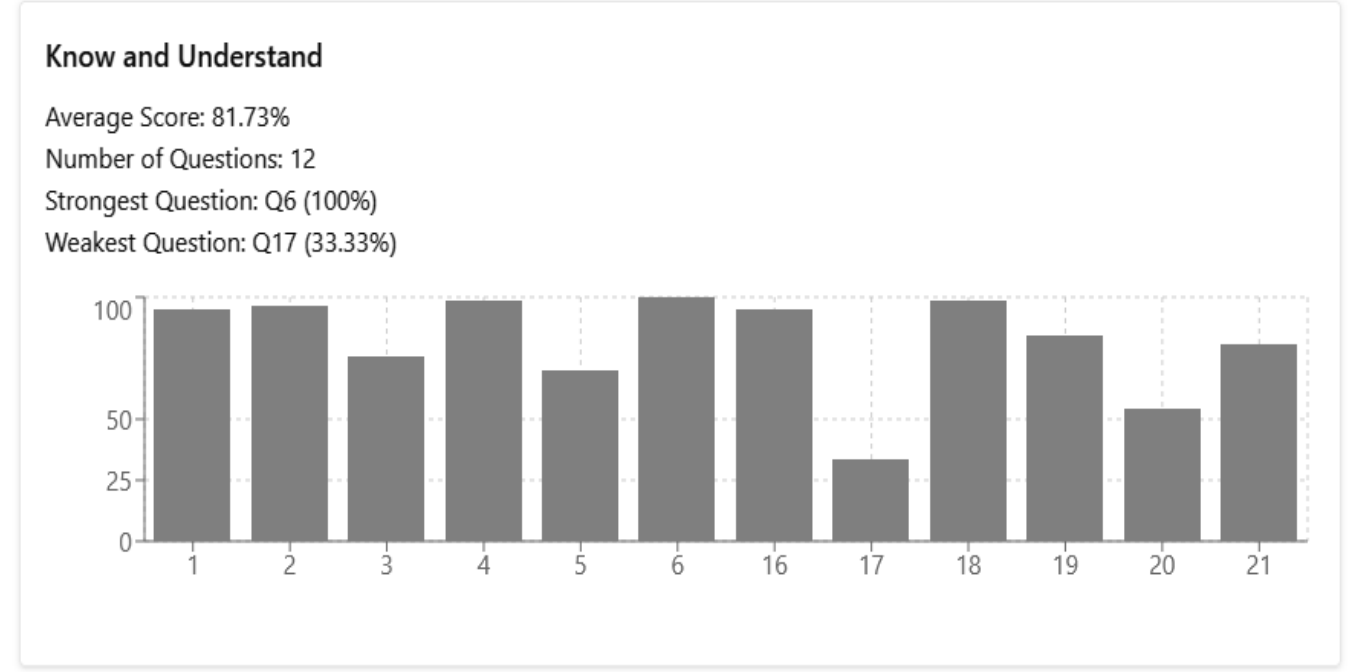

Fig 4: Performance on Know and Understand

## C. *Use and Apply*

Performance in this category was moderate (M = 74.74%; 5 items), suggesting that students had partial knowledge of applying AI to practical tasks. Figure 5 shows the questions in this category arranged by performance. Students could answer questions requiring strong routine applications and near-transfer tasks (Q7–Q9). But the accuracy drops once they start looking at choosing and configuring the correct method. For example, the question tool–task alignment, where many undervalued structured persuasives prompting for a marketing pitch (Q22) and RAG vs. prompting selection, where half misattributed RAG’s advantage (Q23), often generalizing it as “handling new questions” rather than retrieving evidence from a defined corpus.

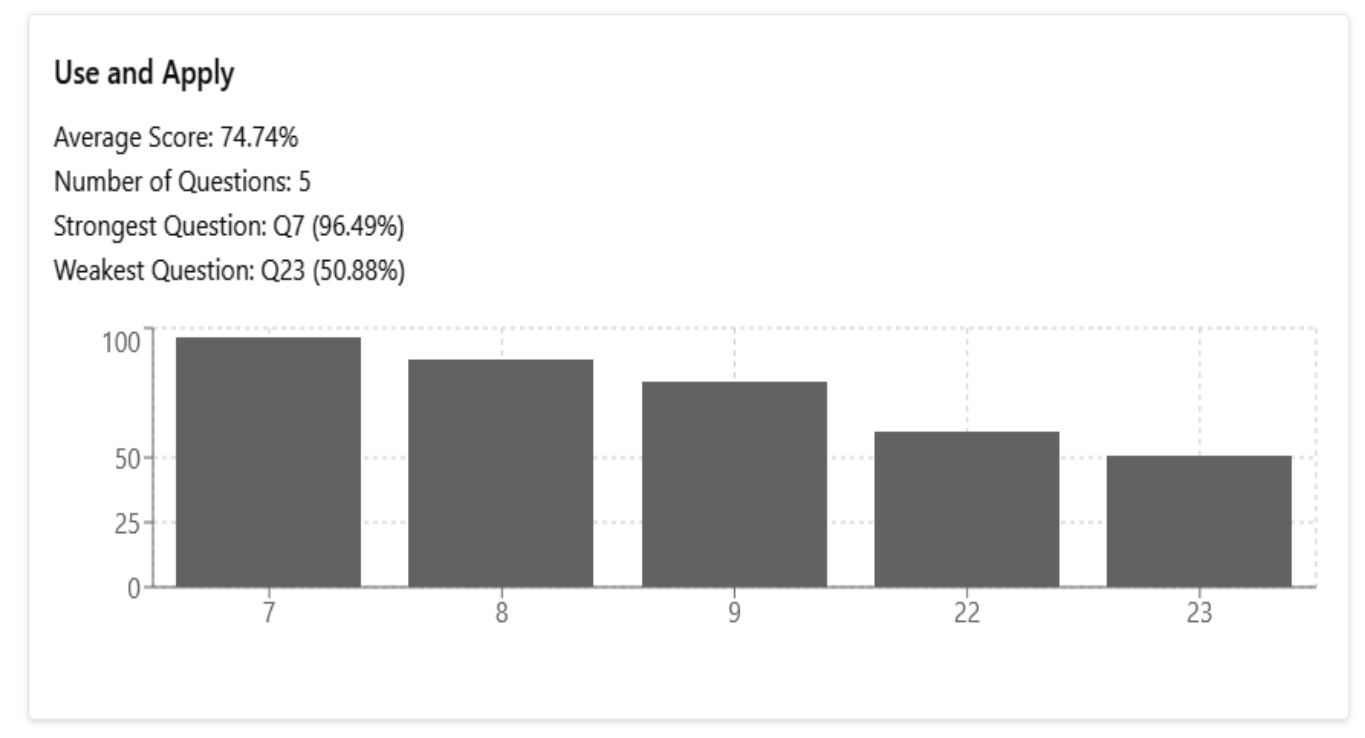

Fig 5: Performance on Use and Apply

## D. *Evalaute and Create*

In this category, students showed mixed results (M=74.94%; 7 items). As shown in Fig. 6 (Q11), about 70% correctly identified the speaker as a device that cannot supply data for computer vision, while about 20% chose a CT scanner, suggesting confusion about modality and task fit. On Q12, around 63% selected more training data as the best lever to improve an underperforming AI system. In comparison, around 30% opted to increase the learning rate, suggesting a bias toward algorithmic tweaks over data-centric improvements.

These misconceptions about sensor relevance and primary data quality can be addressed through brief labs that emphasize modality alignment, dataset curation, and evidence-based error analysis.

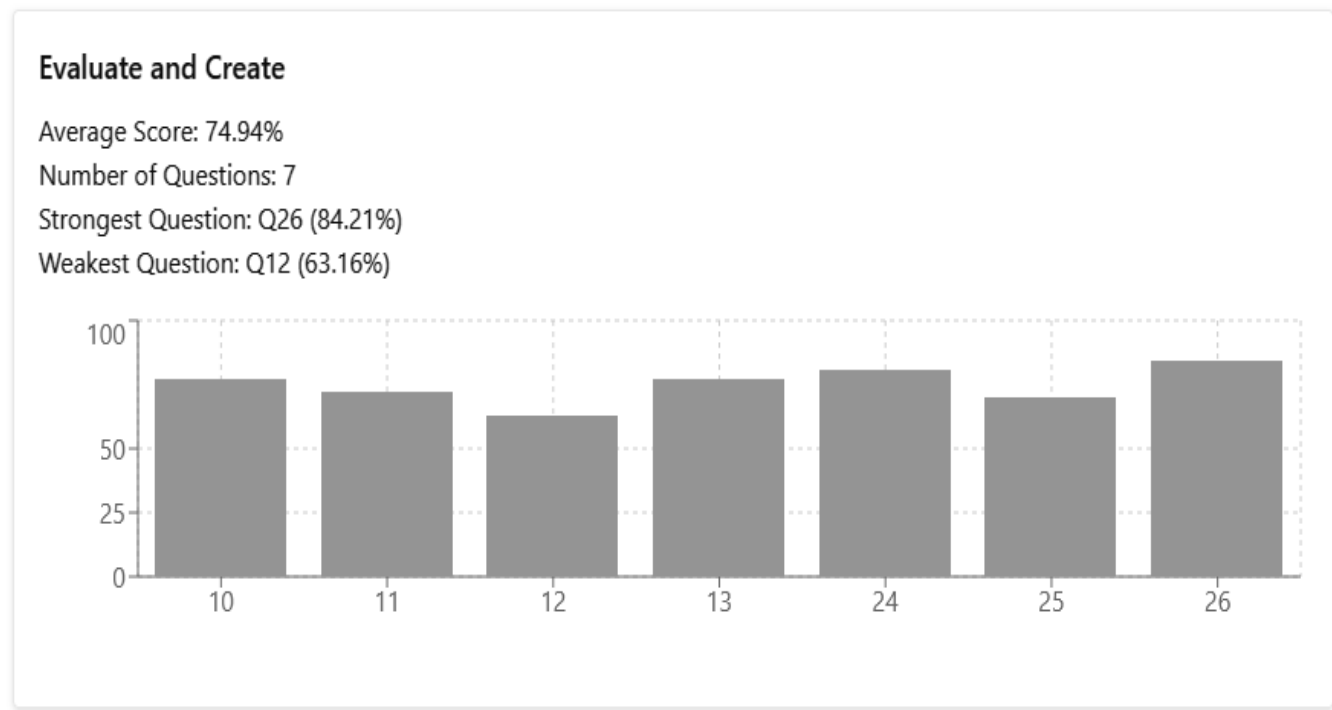

Fig 6: Performance on Evaluate and Create

## *E. Ethics*

Ethics knowledge was strong (M = 80.99%; 6 items), with very high scores on identifying unethical developer behavior (Q14: 98.25%) and on sound judgment regarding explainability concerns in clinical settings (Q15: 84.21%). Most students correctly flagged copyright risk in using AI-generated images commercially (Q27; 85.96%) and showed solid awareness of privacy issues when sharing personal data with cloud-based tools (Q29; 78.95%). The summary of this category is shown in Fig. 7. The question related to the policy tradeoff, however, showed a performance dip, with 64.91% choosing to restrict generative AI outputs to mitigate harm (Q28), indicating confusion around safety and openness. Remarkably, when asked about AI in hiring, 73.68% pinpointed historical data bias as the main fairness risk (Q30), though a minority focused instead on formatting or language-handling errors. In summary, students pinpointed major ethical issues such as bias, copyright, privacy, and transparency, but had difficulty with governance choices where value was at stake.

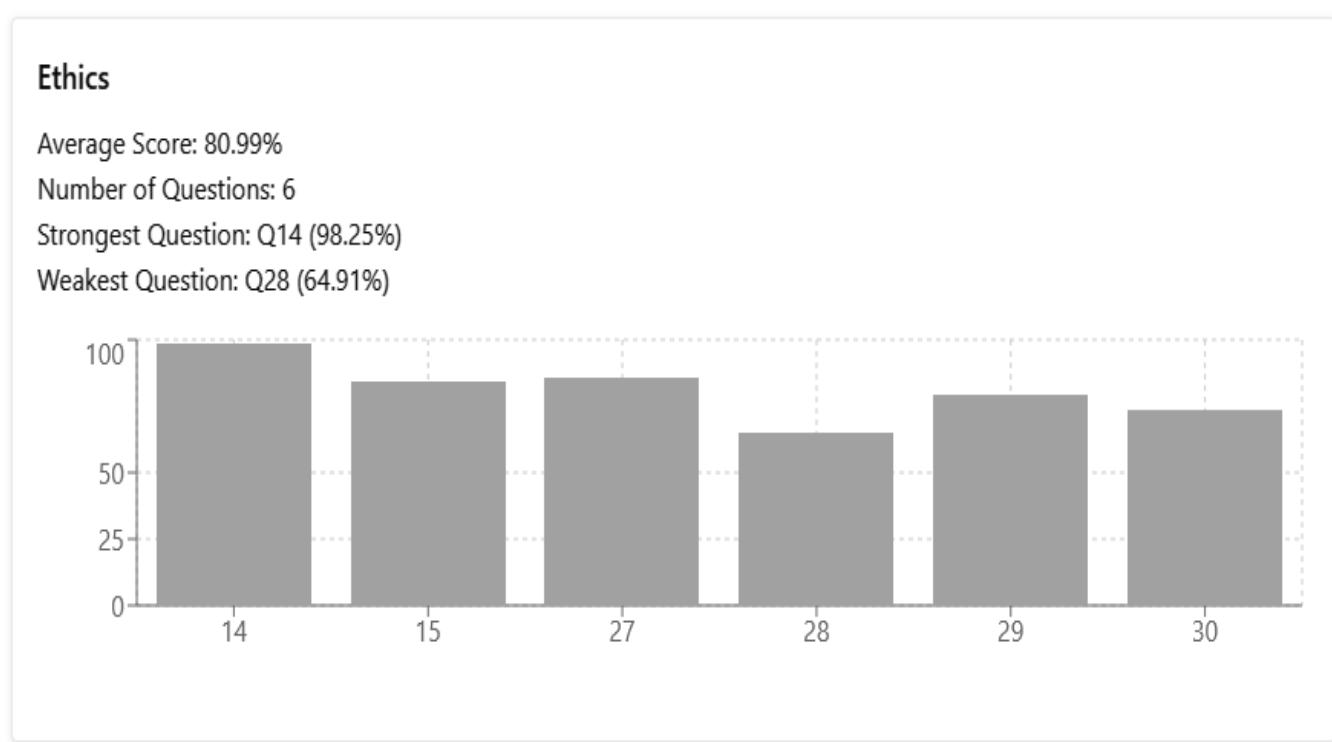

Fig 7: Performance on Ethics

# VI. Discussion

In this paper, we present findings from the evaluation of a knowledge assessment survey designed to evaluate students' understanding of AI and GenAI. The purpose of curating this survey was to create something that could be used both as a formative and diagnostic instrument or as a summative evaluation. The knowledge survey consists of 30 items aligned with four aspects of AI literacy outlined in the literature: know and understand, use and apply, evaluate and create, and navigate ethically. We recommend that, if readers plan to use this survey for summative assessment at the end of the course, they add a few more items directly related to the course content and if the course is related to a domain, the use of AI and GenAI in that domain. Our survey is more general purpose and can be tailored to specific disciplinary needs.

We tested the survey with undergraduate students studying IT. We found that overall, students exhibited a good understanding of where these technologies are implemented and used, but lacked an understanding of how applications that use them work. Specifically, they lacked understanding of how LLMs work and of other concepts such as RAG. The disparity between surface-level awareness of AI systems and technical understanding may explain the lower Cronbach's alpha score. This disparity also aligns with the broader AI landscape, where users have prompting skills but lack the computational foundations of AI systems.

In terms of instruction, when used a formative or diagnostic tool it can assist faculty in design of curriculum. Developing students' understanding of AI as future engineers who may design, implement, or critically evaluate systems is where introductory curricula should be focused. For students in technical domains, it will be necessary for them to move beyond just users to a deeper comprehension of architectural principles. This survey can help identify knowledge gaps. For students in other disciplines, it can assist with getting an understanding of their social or contextual knowledge and their ethical leaning. Finally, this survey, or a variation can be added also to the overall curriculum to assess students' knowledge of AI and GenAI longitudinally [16-17]. Given the increased used of GenAI at the primary or K12 level, it is quite feasible that incoming students would have a higher level of knowledge about AI and therefore the overall curriculum will need constant revisions.

Here are two concrete examples of how we are using the survey in our instruction. The authors teach undergraduate courses one directed at freshmen engineering students and the other at students who transfer from community college. Although the course content varies, one of the objectives is to introduce students to new and emerging technologies but in addition to technical content, emphasize the socio-technical aspects as well. In one class, the survey is used before the final project which is a case study analyzing autonomous vehicles (AV). In this case, specific questions about AV are added to the survey. In the other class, the survey is given at the start of the semester to gauge students' overall knowledge and then the course content is slightly modified to touch upon areas that are seemed to be deficient. In both instances, a pre/post approach is taken to assess if students' knowledge of AI and GenAI has improved over time.

## VII. Limitations and Future Work

There are several limitations of this study that we plan to address in future studies. First, this survey was conducted at a single university with a single cohort of students. Therefore, sampling bias can affect the responses. As a next step, we plan to scale up the implementation to other student populations, including a diverse set of disciplines and institutions. We also plan to expand data collection to non-student populations. Second, although we have used validated survey items to curate this single instrument, it can benefit from future validation studies. Finally, given the rapid pace of change in AI knowledge, it will be prudent to revisit the survey and add or delete items to keep it current.

## VIII. Conclusion

In this paper, we present an assessment instrument for evaluating AI literacy. The instrument combines items from three separate validated instruments to create a comprehensive assessment that can be used with students and other user populations. Unlike other self-reported assessments that target a specific instructional approach or attitudes, this instrument assesses actual content knowledge. We found that the instrument is both comprehensive and discerning; it covers a broad spectrum of AI knowledge but is also useful for identifying misconceptions students might hold about specific aspects of how AI works. Future work will scale up the implementation of the instrument to larger student populations and other user groups.

## Acknowledgment

This work is partly supported by U.S. NSF Award# 2439459, 2439460, 2319137, 1954556, and USDA/NIFA Award# 2021-67021-35329. Any opinions, findings, and conclusions or recommendations expressed in this material are those of the authors and do not necessarily reflect the views of the funding agencies.